\documentclass{appolb}
\usepackage{graphicx}

\usepackage[
    natbib=true,
    backend=biber,
    style=numeric,
    maxbibnames=4,
    maxcitenames=2]{biblatex}

\usepackage{xcolor}

\begin{document}
\title{Denoising medium resolution stellar spectra with U-Net convolutional neural networks}
\author{
{Bal\'azs P\'al
    \address{Department of Physics of Complex Systems, ELTE E\"otv\"os Lor\'and University, Budapest, Hungary}\\
    \address{Institute for Particle and Nuclear Physics, HUN-REN Wigner Research Centre for Physics, Budapest, Hungary}}
\\[3mm]
{L\'aszló Dobos
    \address{Department of Physics \& Astronomy, The Johns Hopkins University, Maryland, USA}
    \address{Department of Information Systems, ELTE E\"otv\"os Lor\'and University, Budapest, Hungary}}
}
\maketitle
\begin{abstract}
    We investigated the use of a U-Net convolutional neural network for denoising simulated medium-resolution spectroscopic observations of stars. Simulated spectra were generated under realistic observational conditions resembling the Subaru Prime Focus Spectrograph (PFS). We found that our U-Net model effectively captured spectral features, achieving an average relative error of around $1\%$ across a broad range of stellar parameters, despite a limited training set of only $1000$ observations and a relatively short training period. Although U-Net did not reach the performance previously demonstrated by fully-connected denoising autoencoders (DAEs) consisting of dense layers and trained extensively on larger datasets, it outperformed dense networks trained under similarly constrained conditions. These results indicate that the U-Net architecture offers rapid, robust feature learning and may be particularly advantageous in scenarios involving initial denoising, subsequently refined by more accurate, but otherwise slower deep-learning models.
\end{abstract}

\section{Introduction}
Spectroscopic surveys of stars often face substantial variations in data quality due to differences of $6$-$7$ magnitudes in apparent brightness, resulting in widely varying signal-to-noise ratios (S/N). Faint, low S/N spectra challenge conventional analysis techniques, such as continuum normalization and template fitting, often leading to non-converging fits or erroneous parameter estimates. Overcoming these limitations is critical for Galactic Archaeology (GA), which relies on spectroscopy to reconstruct the chemical and dynamical history of the Milky Way and nearby galaxies. New survey instruments (e.g., DESI \citep{2022AJ....164..207D, 2023ApJ...944....1D} or Subaru PFS \citep{2014SPIE.9147E..0TS, 2014PASJ...66R...1T}) can now observe very distant and faint stars---including individual red giants at the distance of the Andromeda Galaxy \citep{2023ApJ...944....1D}---offering unprecedented opportunities to study small-scale dark matter dynamics and the early formation of galaxies \citep{2014PASJ...66R...1T}.

To address these observational challenges, machine-learning techniques, such as denoising autoencoders (DAEs) with a dense topology of fully-connected layers have been successfully employed, demonstrating the capability to recover clean stellar spectra from noisy observations at a S/N as low as $10$ with approximately $1\%$ average accuracy \citep{2024AN....34540049P}. Trained on simulated data, DAEs quickly reconstruct essentially noise-free stellar spectra from noisy inputs, enabling near-real-time denoising. This rapid inference capability allows DAEs to be integrated into observation pipelines as immediate on-the-fly tools or as initial approximation steps to enhance the convergence and accuracy of traditional spectroscopic analyses.

In this study, we explore an alternative deep-learning architecture for denoising stellar spectroscopic observations: a U-Net convolutional neural network. The U-Net architecture---originally developed for image segmentation \citep{2015arXiv150504597R}---employs an encoder-decoder structure with skip connections, potentially capturing broad continuum features and sharp spectral lines at the same time more effectively than the simpler fully-connected DAE used in our previous study. We demonstrate that a U-Net model quickly learns robust spectral features early in the training process and achieves reasonable denoising performance even with a very limited training set (e.g. $1000$ observations in fewer than $50$ epochs). However, in the long term, it does not match the performance of dense, fully-connected autoencoders. These results suggest that U-Net may be advantageous in knowledge distillation scenarios, where stellar spectroscopic observations are initially denoised using a U-Net model and subsequently refined using a dense autoencoder.

\section{Simulating stellar spectroscopic observations} \label{sec:synthesis}
To train our neural network model, we generated simulated stellar spectra using the BOSZ synthetic spectrum grid \citep{2017AJ....153..234B}. Observational conditions (seeing, target zenith angle, moon illumination) and instrument parameters were selected to closely resemble typical observing conditions at Mauna Kea Observatory using the Subaru Prime Focus Spectrograph (PFS) in medium-resolution mode ($R \approx 5000$, spectral range between $710$-$885$ nm).

Fundamental stellar parameters ($T_\mathrm{eff}$, $\log g$, metallicity, $\alpha$-element abundances, Doppler shift, magnitude, extinction) were uniformly sampled across broad intervals, covering stellar types M to G. We also included cooler stars to increase spectral variability, to provide a more robust test of the denoising capabilities. Some parameter combinations were non-physical but were intentionally kept to further enhance model robustness.

Each simulated spectrum was produced in two main steps: first, interpolation of the BOSZ grid to randomly selected stellar parameters, applying sub-pixel Doppler shifts; second, application of observational effects including interstellar extinction, instrumental broadening, detector pixelization, and sky noise. The Subaru PFS Exposure Time Calculator \citep{2012arXiv1204.5151H} was adapted and precomputed to efficiently simulate large spectral datasets, ensuring realistic noise modeling.

During training, we further augmented our data by generating new photon-noise realizations for spectra in each epoch and introducing wavelength-dependent flux calibration distortions (up to $2~\%$) to simulate systematic calibration errors. Finally, simulated spectra were normalized to a median flux of $0.5$ within $6500$-$9500$~\AA.

\section{Training}
We implemented a U-Net convolutional neural network to denoise stellar spectroscopic observations, closely following the original U-Net architecture. Our network consists of four encoder-decoder levels, each comprising convolutional blocks with filter counts of $32$, $64$, $128$, and $256$ at successive depths, complemented by max-pooling operations and a bottleneck layer between the encoder and decoder.

The training procedure closely mirrored that of our previous study. However, in this study, both the training and validation datasets contained $1{,}000$ synthetic spectra each, which were augmented at each epoch as detailed in Sec.~\ref{sec:synthesis}. We optimized the U-Net using a stochastic gradient descent optimizer, minimizing the mean squared error (MSE) loss function over $50$ epochs. Batch normalization and \textsc{ReLU} activations were applied in all layers, except the output layer, where we employed a \textsc{Sigmoid} activation. Additionally, we applied a dropout rate of $0.3$ exclusively in the bottleneck layer to mitigate overfitting. The model was trained on a single NVIDIA V100 GPU, with a batch size of $100$ and an initial learning rate of $0.1$, which was reduced by a factor of $10$ after $20$ epochs.

We did not continue training the model beyond $50$ epochs, as we observed that the training and validation losses consistently plateaued after this point.

\section{Results}
We characterize the denoising performance by evaluating the relative errors across the entire validation dataset, comparing the denoised spectra to the original noiseless spectra. Specifically, we calculate the pixel-wise relative error as:

\begin{equation}
    | R_\lambda | = \frac{|D_\lambda[\hat F_\lambda] - F_\lambda|}{F_\lambda}\,,
\end{equation}
where $D_\lambda[\hat F_\lambda]$ denotes the denoised spectrum, $\hat F_\lambda$ the noisy input, and $F_\lambda$ the noiseless simulated spectrum.

\begin{figure}[htb]
    \centerline{%
        \includegraphics[width=12.5cm]{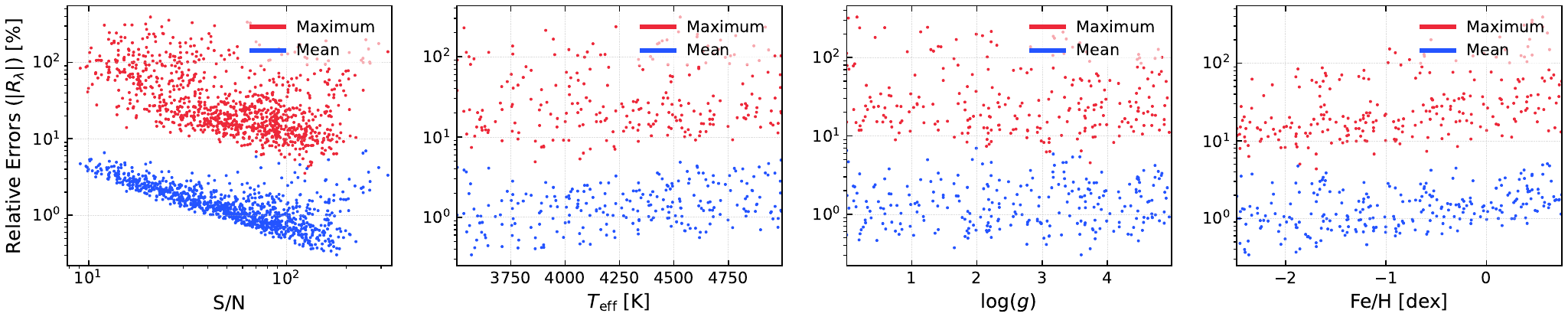}}
    \caption{The pixel-wise average and maximum of the relative errors from denoising, evaluated across the entire validation dataset, are shown as a function of relevant parameters such as signal-to-noise ratio (S/N), effective temperature ($T_\mathrm{eff}$), stellar surface gravity ($\log g$), and metallicity ($[\mathrm{Fe}/\mathrm{H}]$).}
    \label{fig:errors}
\end{figure}

In Fig.~\ref{fig:errors}, we present the pixel-wise average and maximum of these relative errors as a function of four key stellar parameters: signal-to-noise ratio (S/N), effective temperature ($T_\mathrm{eff}$), stellar surface gravity ($\log g$), and metallicity ($[\mathrm{Fe}/\mathrm{H}]$). As expected, the denoising accuracy improves with increasing S/N, showing a pronounced degradation in performance at lower S/N values. The performance is relatively stable across the investigated temperature, gravity and metallicity ranges, with errors slightly ramping up at higher temperatures, although we expected the opposite, as spectra of cooler stars such as M dwarfs are usually more complex. Similarly, errors increase significantly at higher metallicities, reflecting the richer spectral line structure in metal-rich stars.

\section{Conclusions}
We demonstrated that the U-Net architecture is effective at denoising stellar spectroscopic observations, achieving average relative errors around $1\%$ across a wide range of fundamental stellar parameters and signal-to-noise ratios, before it reaches a knowledge plateau. While this level of performance is notable given the limited training dataset and relatively short training duration, it remains approximately an order of magnitude higher than what was achieved previously using denoising autoencoders (DAEs) for a dataset $100$ times larger and trained for $7500$ epochs, but an order of magnitude lower than the performance of DAEs trained on the small dataset for just $50$ epochs.

Our results suggest that although the U-Net does not reach the same precision as dense DAEs in the long term, it offers rapid feature learning and robustness even with limited data. Consequently, U-Net architectures may be particularly beneficial for knowledge distillation approaches, providing efficient initial denoising that can be subsequently refined by more specialized autoencoder models.

\section*{Acknowledgments}
We would like to thank Istv\'an Csabai and Gergely G\'abor Barnaf\"oldi for their valuable comments and suggestions. This work is supported by the generosity of Eric and Wendy Schmidt, by recommendation of the Schmidt Futures program, the Ministry of Innovation and Technology NRDI Office grants OTKA NN 147550, the KDP-2021 program of the Ministry of Innovation and Technology from the source of the NRDI fund and by the the European Union project RRF-2.3.1-21-2022-00004 within the framework of the MILAB Artificial Intelligence National Laboratory.

The simulation of all stellar spectroscopic observations used in this research, and the training of the DAE networks were completed on the Elephant and Volta servers at the Department of Physics \& Astronomy of the Johns Hopkins University. Analysis was carried out on the Volta server, as well as on the Ampere A100 GPU server of the Wigner Scientific Computing Laboratory (WSCLAB) at the HUN-REN Wigner Research Centre for Physics, Hungary.

\printbibliography

\end{document}